\documentclass[12pt]{article}
\usepackage{graphicx}
\begin{document}
\title{\bf {An algebraic approach to the study of 
weakly excited states for a condensate in a ring geometry
}}

\bigskip
\author{
P. Buonsante, R. Franco and V. Penna
\\
\\
{Dipartimento di Fisica and U.d.R. I.N.F.M., 
Politecnico di Torino} 
\\
{C.so Duca degli Abruzzi 24, I-10129 Torino, Italia}
}
%

\maketitle
\begin{abstract}
We determine the low-energy spectrum and the eigenstates for
a two-bosonic mode nonlinear model by applying the In\"{o}n\"{u}-Wigner
contraction method to the Hamiltonian algebra. 
This model is known to well represent a Bose-Einstein condensate
rotating in a thin torus endowed with two angular-momentum modes as
well as a condensate in a double-well potential characterized by 
two space modes.
We consider such a model in the presence of both an attractive and 
a repulsive boson interaction and investigate regimes corresponding
to different values of the inter-mode tunneling parameter.
We show that the results ensuing from our approach are in
many cases extremely satisfactory. To this end we compare
our results with the ground state obtained both numerically
and within a standard semiclassical approximation based on
su(2) coherent states.
\end{abstract}




\section{Introduction}
\label{S:level1}

The dynamics of a bosonic fluid
rotating within a thin torus and, particularly, the study of the
properties relevant to its weakly-excited states have received recently
a large attention \cite{R}-\cite{KSU2} due to
the rich phenomenology that characterizes such a system. 
For example, the quantization of fluid circulation 
is shown \cite{UL} to disappear whenever the physical parameters cause the
hybridization of
condensate ground state over different angular momentum (AM) states.
A similar effect is found in the mean-field dynamics of the condensate
wavefunction on a circle \cite{KSU1}, where the circulation
loses its quantized character when the system is in the soliton regime.
The rotating fluid exhibits low-energy
AM quantum states (corresponding to the presence of plateaus of quantized circulation) that determine the hybridization effect by
a suitable tuning of the model interaction parameters \cite{UL}. 
In the simplest possible case,
the model exhibits two momentum (bosonic) modes associated to two AM
states (the ground-state and the first excited state) of the fluid.
An almost identical model \cite{milb}-\cite{tonel}
has been studied thoroughly in the recent
years within Bose-Einstein condensates (BEC) physics, where a condensate
is distributed in two potential wells that exchange bosons via
tunneling effect. The two-well model $H= U(n_0^2 + n_1^2) - \Delta (n_0- n_1)/2
-V_0 (a_0 a^+_1 + a_1 a^+_0)$, where $a_0$, $a_1$ are bosonic space-modes and
$n_i =  a^+_i a_i$, displays hybridized states when the well-depth imbalance vanishes ($\Delta =0$). 

For both models the energy regime of interest is that
corresponding to the ground-state or to weakly excited states.
In this respect, many authors have tried to develop
approximation schemes able to provide a satisfactory analytical description
of the low energy spectrum and of its states. The nonlinear character
of the model Hamiltonian entails a difficult diagonalization process
unless one resorts to numerical calculations. In this case
the exact form of the spectrum is obtained quite easily. However,
for N-well systems such as condensate arrays described by 
the Bose-Hubbard model, Josephson-junction arrays and, in general,
N-mode bosonic systems \cite{Sol}, \cite{SolPe}, 
the exact diagonalization requires a
computational effort rapidly increasing with N.
This motivates the interest in developing effective, analytical
approximation methods able to solve the diagonalization problem.

The present work has been inspired by papers \cite{Leg} and
\cite{UL} where, among the variuos issues considered, the structure
of the ground state of a ring condensate (within a two-AM-mode
approximation of the bosonic quantum field) has been studied.
As to the closely related two-well boson model, the same problem
has been investigated in \cite{kostrun} within the hermitian phase
operator method.
In order to obtain a satisfactory description 
of the system ground state as well as of the weakly-excited states 
for the two-mode model, we implement, in the present paper,
an algebraic approach based on
the {\it In\"on\"{u}-Wigner contraction} method \cite{InWig}. 
This method allows one to simplify the algebraic structure of
the Hamiltonian reducing the latter in a form apt to perform a
completely analytic derivation of its spectrum. A well defined limiting
procedure, mapping the original Hamiltonian generating
algebra to a simpler algebra, often succeeds in reducing the
nonlinear terms to a tractable form.
These terms, originated by the boson-boson interaction and thus
occurring in any model inherent in BEC dynamics, are known to
make the Hamiltonian diagonalization a hard task. 
Such a technique and the effect of simplifying the algebraic structure
of model Hamiltonians, has found a wide application in many fields 
of theoretical physics. It is well illustrated, e. g., in reference \cite{rowe}
where it is applied to study collective phenomena in nuclear models.

The contraction-method approach (CMA) --namely
the contraction procedure and the ensuing approximation of 
weakly excited states-- works well for the spectrum sectors where 
the energy levels are close to the minima and the
maxima of the classical Hamiltonian and thus seems suitable
for studying the low-energy regime of two-mode nonlinear models.  
The results obtained within the CMA in sections \ref{sez2}
and \ref{sez3} will be compared both with the exact spectrum
calculated numerically and with an alternative approch based on the
coherent-state semiclassical appproximation (CSSA) reviewed in section
\ref{sez4}.

We consider $N$ interacting bosons with
mass $m$ whose boson-boson interaction can be
either attractive or repulsive.
These are confined in a narrow annulus whose thickness $2r$
is much smaller than the annulus radius $R$. Bosons are also acted 
by an external potential which causes inter-mode tunneling. 
Particularly, the rotating fluid with attractive interaction can be shown
to be equivalent to the two-well model of repulsive bosons 
introduced previously. 
In the coordinate frame of the potential rotating with angular
velocity $\omega$ and with $z$ axis parallel to total angular momentum
$L_{tot} = {L}_z$, the bosonic-field Hamiltonian reads
%
$$
{\hat H}_{bf} = \int d^3 \, {\bf r} {\hat\psi}^+_{\bf r}
\left[\frac{P^2}{2m}-\omega L_{z}+ V_{ext}({\bf r}) \right]
{\hat\psi}_{\bf r} +
\frac{1}{2} \int \! \! \int d^3 {\bf r} \, d^3 {\bf s} \, 
{\hat\psi}^+_{\bf r} \, {\hat\psi}^+_{\bf s}
 U (|{\bf r}- {\bf s}|) {\hat\psi}_{\bf s} \, {\hat\psi}_{\bf r}
$$
%
where ${\hat\psi}_{\bf r} 
= \hat\psi({\bf r})$ (${\hat\psi}^{+}_{\bf r}$)
is the destruction (creation) boson field operator 
at ${\bf r}$. $V_{ext}$ is the confining potential.
At low temperature, the interaction between dilute
bosons is well represented by the Fermi contact
interaction which entails the standard approximation
$U (|{\bf r}- {\bf s}|) \simeq
({4\pi \hbar^2 a}/{m})\delta(|{\bf r}- {\bf s}|) $, where
$a$ is the $s-$wave scattering length \cite{Leg}. 

\subsection{Two-mode approximation}
\label{sec1.1}

The two-mode approximation involves only the first two states of
AM, with eigenvalue equations $L_{z}\psi_{0}({\bf r})=0$ and
$L_{z}\psi_{1}({\bf r})=\hbar \psi_{1}({\bf r})$. Field
operator ${\hat \psi}({\bf r})$ in the two-mode basis of $L_{z}$ 
is thus written as
${\hat \psi}({\bf r}) \simeq {a}_{0}\psi_{0}({\bf r})+ 
{a}_{1}\psi_{1}({\bf r})$,
where $a_0$, $a_1$ are bosonic operators and the validity of the
two-mode approximation requires the condition $0< \omega < 2\omega _{c}$ 
(greater angular velocities would involve other angular-momentum states).
Within such an approximation~\cite{milb, macz}
and considering a thin torus ($r << R$) 
${\hat H}_{bf}$ reduces to \cite{Leg}
$H = {g} (n^2_0 + n^2_1- n_0 - n_1 + 4n_0 n_1)/{2}
-{\Delta} \hbar n_1/2  -V_0 (a_1^+ a_0 +a_0^+ a_1)$, 
where $n_i= a^+_i a_i$, $\Delta=2\hbar (\omega - \omega_{c} )$, 
while $\omega_{c}= \hbar/(2mR^{2})$, $g= 2\hbar^2 a/(m R \pi r^2)$, and 
$V_0$ are the critical angular frequency, the mean interaction energy
per particle, and the asymmetry of potential $V_{ext}= V_0 (e^{i\theta}
+e^{i\theta})$, respectively. In the Schwinger picture \cite{Steel} 
of algebra su(2) $H$ further simplifies becoming, up to a constant term,
\begin{equation}
H= -g \, J_{3}^{2} -2V_{0} J_{1}-\Delta J_{3}\, ,
\label{Hmod}
\end{equation}
where $J_{3}=({n}_{1}-{n}_{0})/2$, $J_1 = (J_+ +J_-)/2$, $J_2 = (J_+ -J_-)/2i$
and $J_{+} = a^{+}_{1} a_{0}$ , $J_{-} = (J_+)^+$. 
Such generators satisfy the commutators
$[J_r, J_s]= i \epsilon_{rsv} J_v$ ($\epsilon_{rsv}$ is the antisymmetric 
symbol) and commute with the total boson number operator ${n}_{1}+{n}_{0}$ 
($n_i= a^{+}_i a_i$) whose eigenvalue $N$ is connected with the 
su(2)-representation index $J$ by $J=2N$. 
In such a scheme, the AM states are defined by
$$
|J;m\rangle :=|n_{0}\rangle \otimes |n_{1}\rangle ,
\quad n_{1}=J+m, \,\, n_{0}=J-m ,
$$
where the $J_3$-basis states satisfy the eigenvalue equations 
$J_{3} |J;m \rangle =m |J;m \rangle$, and 
$J_{4} |J;m \rangle =J |J;m \rangle$, the index $J$ being 
the eigenvalue of $J_{4}= (n_1+ n_0)/2$. 
The positive (negative) sign of $g$ in model \ref{Hmod} implies that the effective interaction between bosons is repulsive (attractive).
The conditions of weak asymmetry and interaction ensuring the
validity of the model \cite{Leg} are given by 
$|V_{0}|\ll \hbar \omega_{c} $, and $|g|\ll \hbar\omega_{c} $. 
The simple spin form of Hamiltonian \ref{Hmod}
evidences how the attractive model ($g<0$) 
coincides with a (repulsive) two-site Bose-Hubbard Hamiltonian
\cite{Steel, FPZ1} modeling two potential wells of different
depth that share $N=2J$ bosons and exchange them via tunnel effect. 
The $N$-boson physical states can be written as
$
| \psi \rangle = \sum_{m=-J}^{J} X_m |J; m \rangle \, ,
$
while the Schr\"odinger equation $(i\hbar \partial_t -H )|\psi\rangle=0$ 
can be expressed in components as
$$
i\hbar {\dot X}_m = \left ( -g m^{2} - m \Delta \right ) X_m
-V_0 \left [ R^J_{m+1}X_{m+1}  + R^J_m X_{m-1} \right ] ,
$$
once the symbol $R^J_m = [(J+m)(J-m+1)]^{1/2}$ has been defined.
It worth noting that the study the algebraic structure characterizing 
the second-quantized Hamiltonian for a condensate trapped in two potential
wells has received a large attention in the literature. In the seminal
work \cite{Sol} and in reference \cite{SolPe},
in particular, such
Hamiltonian has been shown to reduce, within a standard mean-field approach, 
to the sum of mode Hamiltonians describing
the momentum conservation in the presence of inter-well boson
exchange due to the tunneling. Each mode Hamiltonian is written in
terms of operators $a_k$, $a_{-k}$ ($\pm k$ are the momentum modes)
and can be reformulated as a linear combination of su(1,1) generators.
In model \ref{Hmod} the momentum conservation is explicitly violated
since one of the mode takes into account the fluid rotation. 
This fact entails that the previuos Schwinger realization of algebra
su(2), rather than the algebra su(1,1) connected with the
momentum conservation, characterizes the system.

In our analysis
the dimensionless mean-value per boson of the angular momentum
$\left<l_{z}\right>= \langle L_{z}\rangle /{\hbar N}$ (where the
notation $\langle A\rangle = \langle \psi| A | \psi \rangle $ has
been introduced) represents an important quantity. 
The angular momentum, in fact, expressed as
\begin{equation}
\left<l_{z}\right>=\sum_{m=-J}^{J}\frac{J+m}{2J}|X_m|^{2} =
\left( \frac{1}{2}+\frac{\left<J_{3} \right>}{2J} \right) ,
\label{lz}
\end{equation}
relates the macroscopic behavior of the rotating condensate
to the minimum-energy state properties through the ground-state
components $X_m$.
In the sequel we consider the spectral properties of model \ref{Hmod}
both in the attractive case ($g<0$)
\begin{equation}
H_a = |g| \, J_{3}^{2}-2V_0J_{1}- \Delta J_{3} ,
\label{Hatt}
\end{equation}
and in the repulsive case ($g>0$)
\begin{equation}
H_r = - \left( |g|\, J_{3}^{2}+2V_0J_{1}+ \Delta J_{3}\right) .
\label{Hrep}
\end{equation}
It is worth noting that the study the ground-state properties of the repulsive case is closely related to the study the maximum-energy state for the 
attractive Hamiltonian. In fact, after the 
substitutions $V_0 \rightarrow -V_0$ and $\Delta \rightarrow -\Delta$, 
the repulsive Hamiltonian is identical to the attractive one up to 
a factor $(-1)$. Since these two changes can be effected in a unitary
way by means of transformations
$e^{+i\pi J_3 } J_1 e^{-i\pi J_3 } =-J_1$, and 
$e^{+i\pi J_1 } J_3 e^{-i\pi J_1 } =-J_3$,
respectively, the spectra of $H_r$ and $H_a$ turn out to satisfy the equation 
$spect [H_a (V_0, \Delta)]$ $=-spect[ H_r (-V_0, -\Delta)]$.
Concerning the parameter $\Delta$ of Hamiltonian \ref{Hmod},
we note that the constraint $0 \le \omega \le 2\omega_c$,
implies the inequality $-2 \hbar w_c < \Delta < 2 \hbar w_c $.
The definition of the further parameters 
$\gamma = {J |g| }/{2 \hbar w_c}$, $\tau ={V_0}/{J |g|}$,
allows one to better characterize the regimes of the rotational dynamics
as well as the conditions of validity of the present model.
Parameter $\gamma$ (representing the ratio of the self-interaction energy
per particle to the single-particle energy-level spacing) should satisfy
the inequalities $2 \gamma << J$, $\tau << {1}/{2 \gamma }$,
owing to the conditions $|g|<< \hbar \omega_{c} $ and
$V_0 << \hbar \omega_{c}$, respectively. Both these conditions can be
satisfied if $J=N/2$ is not excessively large \cite{Leg}.
Moreover, parameter $\tau={V_0}/(J|g|)$ allows one to distinguish, 
in both the attractive and repulsive case, three regimes:
\begin{itemize}
\item Fock regime, where $|g|\gg V_0 J$ entails $ \tau \ll {1}/{J^2} $,
\item Josephson regime, where $ V_0 /J \ll |g| \ll V_0 J$ entails
${1}/{J^2} \ll \tau \ll 1 $,
\item Rabi regime, where $|g| \ll V_0/J $ entails $\tau \gg 1 $.
\end{itemize}
We note that the condition of weak asymmetry $|V_0|\ll \hbar
\omega_c$ given by $\tau \ll {1}/{2 \gamma}$ appears to be 
compatible with the first two regimes and with part of the Rabi regime.

\section{The In\"{o}n\"{u} - Wigner contraction in the attractive case}
\label{sez2}

We introduce a simple algebraic approach for studying the 
low-energy spectrum of Hamiltonians \ref{Hatt} and \ref{Hrep} for 
large $J$ whose essence consists in simplifying the nonlinearity
due to the term $J^2_3$. The In\"{o}n\"{u}-Wigner contraction 
\cite{Amico} supplies 
a method for mapping some given algebraic structure in a new one,
as the result of a singular limiting process. The contraction is 
realized by defining a set of new operators $h_i$ as linear 
combinations $h_i= \sigma_i I+ \Sigma_k c_{ik} g_k$ of the generators 
$g_k$ of a given algebra (identified by its commutators 
$[g_r, g_s]= \varepsilon_{rsk} g_k$) and of the identity operator $I$. 
Selecting an appropriate parametrization $c_{ik}(x)$ of the linear-map
coefficients, the contraction enacted by means of the limit
$ x\to 0$ is able to generate the new
algebraic structure $[h_i, h_j]= e_{ijk} h_k$ 
whose structure constants $\{ e_{ijk}\}$ differ from the
original ones $\{\varepsilon_{rsk}\}$.
For the algebra su(2) the contraction of the algebra mapping
is driven by $x=1/\sqrt J$ (with $J\rightarrow \infty$) and generates, in this 
limit, the harmonic oscillator (namely the Heseinberg-Weyl) algebra 
\cite{FP2}. 

The classical study of attractive ($g<0$) Hamiltonian
$H_a = |g| \, J_{3}^{2}-2V_0J_{1}- \Delta J_{3}$
developed in \ref{appB} demonstrates (see formula \ref{min1})
how $J_1 \simeq +J$, $J \gg |J_2|, |J_3| \simeq 0$, at low energies.
This suggests the correct way to implement the contraction scheme.
In the present attractive case we can build the following
transformation
\begin{equation}
h_{1} =J_1-{I}/{x^{2}}, 
\, \,
h_{2} =xJ_2, 
\, \,
h_{3} =x J_3,
\label{contr1}
\end{equation}
where $J_4 = J\,I$. The In\"{o}n\"{u}-Wigner contraction is realized when
such a $x$-dependent transformation is considered in the (singular) limit 
$J =1/x^2 \rightarrow \infty$. In this
case the objects $\{ J_{i} \}$ (with $i=1,2,3,4$), defining
algebra $u(2)$, transform into the new objects $\{ h_{i},I \}$
(with $i=1,2,3$) that satisfy the following commutation relations:
\begin{equation}
[h_2,h_3]=i(x^2 h_1 +I) \to iI, \,\,
[h_1,h_2]=x[J_1,J_2]= ih_3 \, , 
\end{equation}
\begin{equation}
[h_1,h_3]= x[J_1,J_3]= - ih_2 \, , \,\,
[h_i \, ,\, I\, ]= 0 \, .
\end{equation}
In the limit $x={1}/{\sqrt{J}} \to 0$, the latter reproduce the
commutation relations of Weyl-Heisemberg algebra: $[q,p]=i$,
$[n,q]=-ip$, $[n,p]=iq$, $n= (q^2+p^2)/2$, thereby suggesting 
the identifications $h_1 \equiv -n$, $h_2 \equiv -p$, $h_3 \equiv q$.
By combining the latter with definitions \ref{contr1} 
we find that the contraction gives $J_{1}\rightarrow J-n$, 
$J_{2}\rightarrow -\sqrt{J} p$, $J_{3}\rightarrow \sqrt{J} q$.
%
%
%
Correspondingly, Hamiltonian $H_a$ becomes
$H_a =|g|Jq^{2}+2V_0n-2V_0J -\Delta \sqrt{J}q$,
which, by defining
$\Omega= [1+{1}/{\tau}]^{1/2}$, and $Q= q -\chi$ with 
$\chi= {\sqrt{J} \Delta}/{ 2 V_0 \Omega^{2}}$,
and $\tau = V_0/J|g|$, reduces to the form
\begin{equation}
H_a =
V_0 \left [{p^{2}}+ 
\Omega^{2} Q^{2}-2J -\frac{J \Delta^{2}}{4V_0^2 \Omega^{2}} \right ] \, .
\label{aHO}
\end{equation}
Since ${p^{2}}+\Omega^{2} Q^{2} = 2\Omega (n+1/2)$ is diagonalized
by the harmonic-oscillator eigenstates 
$\Psi_n (Q)= \langle Q|E_{n} \rangle =N_{n} e^{-\Omega Q^{2}/{2} }
H_{n} \left ( {\sqrt \Omega Q} \right ) $,
the eigenvalues of Hamiltonian \ref{aHO} are found to be
\begin{equation}
E_n= 
V_0 \left [2\Omega (n+1/2) -2J -\frac{J \Delta^{2}}{4\Omega^{2} V^2_0} 
\right ] \, .
\label{EAval}
\end{equation}
The corresponding eigenvalue equation 
$H_a |E_{n}\rangle = E_{n} |E_{n}\rangle $ in the
$J_3$ basis, where $|E_{n}\rangle = \Sigma_m X_n(m) |J,m \rangle$, 
can be written as $\Sigma_m (H_a)_{\ell m} X_n(m) = E_{n} X_n(\ell) $
with $(H_a)_{\ell m} = \langle J,\ell |H_a|J,m \rangle$.
In the limit $J >>1$, equation $J_3 |J,m \rangle = m|J,m \rangle $ is replaced by
$q|J,m \rangle = (m/{\sqrt J}) |J,m \rangle $. Therefore
the eigenvalue $m/\sqrt J$ can be seen as a 
continuous variable which naturally identifies with the variable 
$q \simeq J_3/\sqrt J$ used within the approximation scheme just discussed.
The component version of the eigenvalue equation for $H_a$  
then reduces (see reference \cite{FP2} for details) to the 
equation $H_a(Q,p) \Psi_n (Q)= E_n \Psi_n (Q)$ solved above. 
Components $X_m (E_{n})$ thus appear to be given by
$X_m (E_{n}) = \Psi_n (Q)$ that entail the explicit expression for the eigenstates 
\begin{equation}
|E_{n} \rangle = \Sigma_m X_m (E_n ) |J; m \rangle ,
\quad X_m (E_n ) = N_{n}H_{n}({\sqrt \Omega Q}) e^{-{\Omega Q^{2}}/{2}}
\label{EEAtlarge}
\end{equation}
with $Q= {m}/{\sqrt J} -\chi$.
The normalization constants $N_n$ are determined through the condition
$\langle E_{n}|E_{n} \rangle = 1$ implying that
\begin{equation}
\! 1= \sum^J_{m=-J} X^2_m(E_n) 
\simeq
\int^{\infty}_{-\infty} dq \frac{N^2_n }{\sqrt J} \,  
H^2_n \left [ {\sqrt \Omega}(q-\chi) \right ] \,
e^{-\Omega(q-\chi)^2} \, ,
\label{norm1}
\end{equation}
where $\pm J$ has been replaced with $\pm \infty$. Such an approximation is
acceptable until the condition 
\begin{equation}
|\chi| < {\sqrt J}- \sqrt {2n/ \Omega }
\label{Herm}
\end{equation}
--evinced from the interval containing the Hermite-polynomial zeros--
is fulfilled. Excluding the case
$\tau \gg 1$, this condition is always valid provided $n<<J$. 
Thus constants $N_n$ are given by
$N_n= [(J \Omega)^{1/2} / ( \pi^{1/2} 2^n  n!)]^{1/2} $.

Another important check concerns the possibility of considering $m/\sqrt J$ 
as a continuous variable. The characteristic scale is established by the 
gaussian deviation $\sqrt {2/ \Omega}$ which must be compared with the
smallest variation $1/\sqrt J$ of $q$. The resulting condition 
$1/\sqrt J < \sqrt {2/ \Omega}$ can be written as
$$
1 < \frac{ 2J}{\Omega}  = 2J \left [ \frac{V_0}{V_0 + J|g|}  \right ]^{1/2} =
2J \left [ \frac{\tau}{\tau + 1}  \right ]^{1/2}.
$$
While in the Rabi and Josephson regimes ($1/J^2 <<\tau$)
the latter is fully satisfied,
in the Fock regime, where $\tau << 1/J^2$,
such condition is violated.
We notice that for $\tau \simeq 1/J^2$
(namely $|g|\approx JV_0$) a unique component $X_m$ appears to
contribute to states $|E_n \rangle$ since the gaussian amplitude becomes
very small. For example, in the case of the ground-state one has
\begin{equation}
|E_{0}\rangle = \Sigma_m 
N_{0} 
e^{-\frac{\Omega}{2} 
\left ( \frac{m}{{\sqrt J}} -\chi \right )^{2} } |J,m \rangle
\simeq  N_{0} e^{- \frac{\Omega}{2} 
\left ( \frac{m_*}{{\sqrt J}} -\chi \right )^{2} } |J,m_* \rangle \, ,
\label{Fgstate1}
\end{equation}
where $m_*$ is the integer closest to ${\sqrt J }\chi \simeq \Delta/2|g|$.
Nevertheless,
in the special case when $\Delta/2|g| = m_* +1/2$, the two states $|J,m_* \rangle$
and $|J,m_*+1 \rangle$ equally contribute to $|E_{0}\rangle$ which is given by
\begin{equation}
|E_{0}\rangle
\simeq  N_{0} e^{-\frac{\Omega}{8J} } (|J,m_* \rangle +|J, m_* +1 \rangle ). 
\label{Fgstate2}
\end{equation}
To summarize, we note how the ground-state $|E_{0}\rangle$
is essentially formed by a unique component corresponding to $|J,m_* \rangle$
in the whole parameter range $ m_*-1/2 < \Delta/2|g| < m_*+1/2$. The resonance
of the system between two equivalent states crops up whenever $\Delta/|g|$
assumes integer values given by $\Delta/|g| \equiv 2m +1$ with $-J \le m \le J$.
Such condition can be implemented by varying $\Delta$ with $|g|=const$ thus
leaving $\Omega$ unchanged.

\begin{figure}
\begin{center}
\begin{tabular}{cc}
\includegraphics[width=6.5cm]{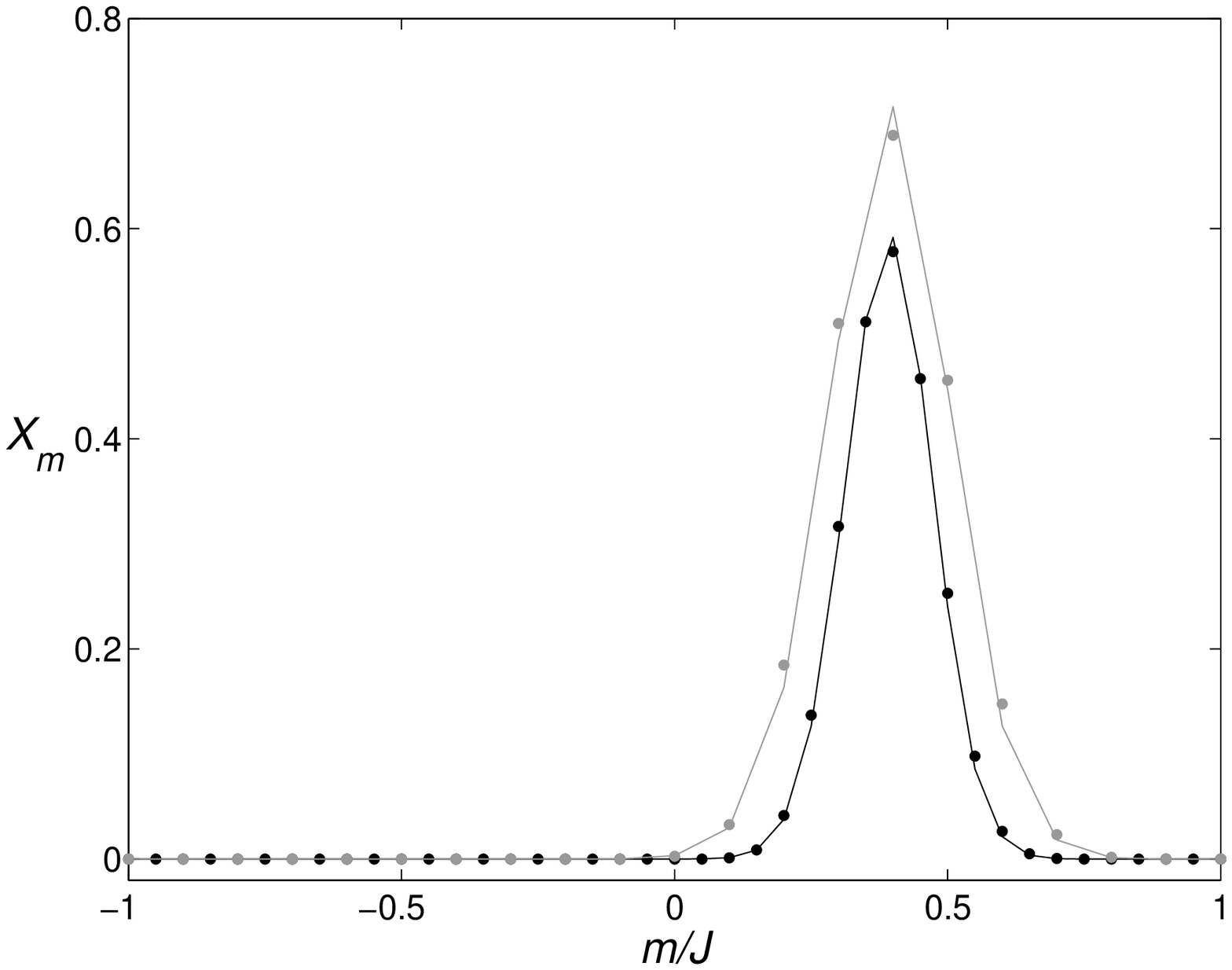}&
\includegraphics[width=6.5cm]{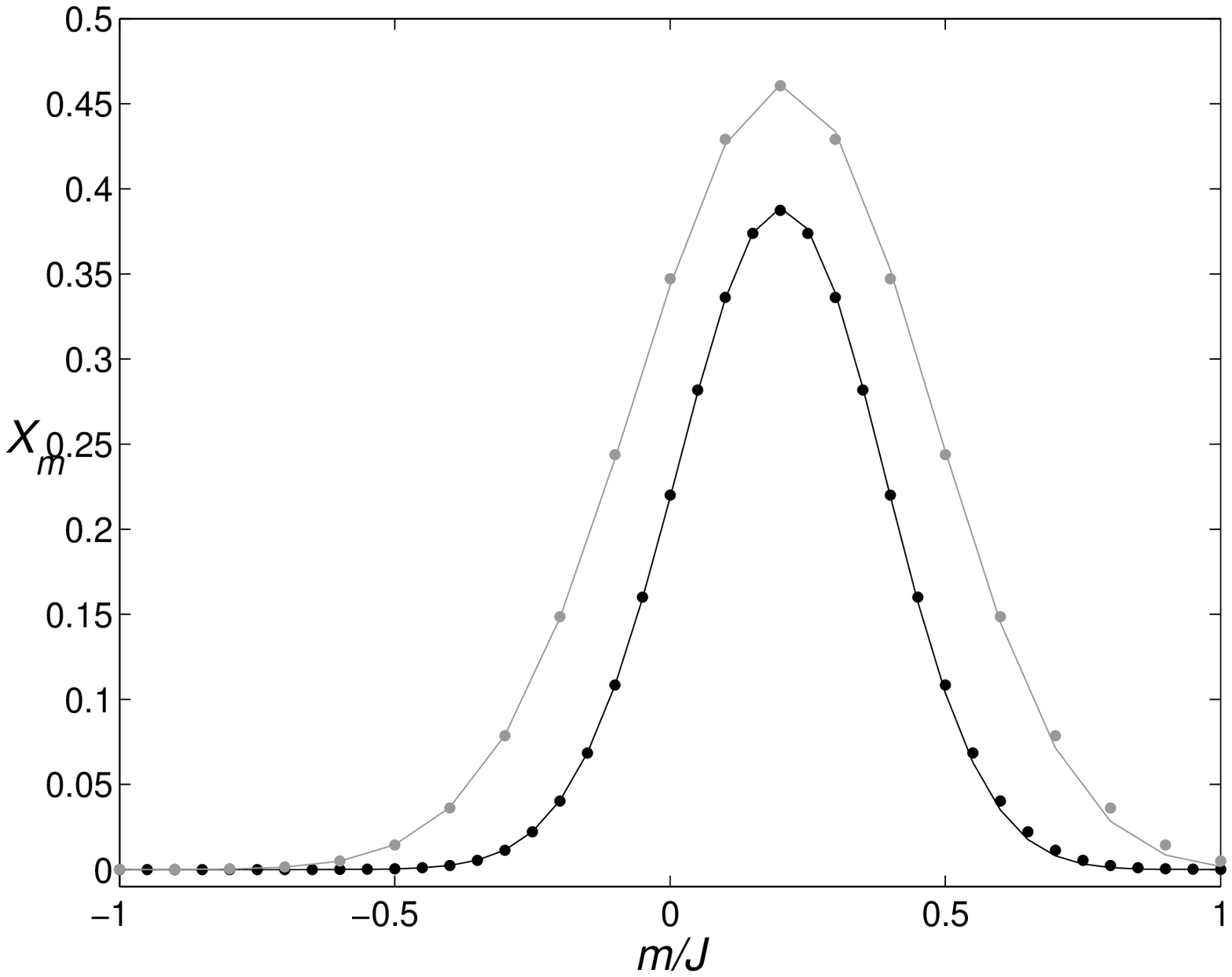}
\end{tabular}
\caption{\label{figura12} In both panels,
grey (dark) diamonds describe the ground-state components $X_m$,
obtained from formula \ref{EEAtlarge}, for
$N=20$ ($N=40$) within the contraction-method approach (CMA).
The edges of the grey/dark piecewise linear curves represent
components $X_m$'s calculated numerically.
{\bf Left panel}: 
Josephson regime in the attractive case with $\tau=0.02$, $\nu =0.8$. 
CMA components $X_m$'s and $X_m$'s calculated numerically
are almost indistinguishable. 
{\bf Right panel}: 
Attractive case with $\tau=1.0$ (transition point
from Josephson to Rabi regime) and $\nu =0.8$.
No difference is visible between CMA $X_m$'s 
and $X_m$'s calculated numerically.
}
\end{center}
\end{figure}

\subsection{Comparison of different regimes} 
For $\tau > 1/J^2$ (Rabi and Josephson regimes), 
one easily calculates the dimensionless mean  AM per boson 
$\langle l_z \rangle $
based on state $|E_{0}\rangle$, as given by formula \ref{EEAtlarge}, 
and exploiting the normalization integral \ref{norm1}. Recalling that
$\langle J_3 \rangle =\sum^J_{m=-J} m X^2_m(E_0)$, one finds
\begin{equation}
\! \! 
\langle l_z \rangle = 
\frac{1}{2} \left (1 + \frac{\langle J_3 \rangle}{J} \right ) 
\simeq \frac{1}{2}+ \frac{\tau \Delta }{4 V_0 (1 + \tau)}
\, , \,\,\,
\langle J_3 \rangle \simeq {\sqrt J} \chi 
=
\frac{{J} \tau \Delta  }{ 2 V_0 (1 + \tau)},
\label{J3}
\end{equation}
%
%
where $\langle J_3 \rangle$ matches exactly formula \ref{minR}
obtained in the classical study of the attractive model.
This result cannot be used in the Fock regime where
the ground state has, at most, either one or two dominating components.
In the other two regimes, the second of equations \ref{J3} entails the further
consistence condition
\begin{equation}
-1 \le \langle J_3 \rangle/J = 
{ \tau \Delta  }/{ [2 V_0 (1 + \tau)]} \le +1 \, ,
\label{J3vin}
\end{equation}
which has to be verified in each regime. In view of the condition 
$|\langle J_3 \rangle|<< J$ required to implement
the contraction procedure, formula \ref{J3vin} should be imposed
in the stronger version $ | \tau \Delta /[ 2 V_0 (1 + \tau)]| << 1$.
However, the numerical (exact) determination of the ground state for various choices of parameters reveals that our approximate procedure works well also in the case when $| \tau \Delta /[ 2 V_0 (1 + \tau)]|$ is not particularly small.
\medskip

\noindent
\textbf{Fock regime}.
The main feature of this case ($\tau \ll {1}/{J^2}$)
is that the mean dimensionless AM per boson is a step function of $\Delta$
(as to this well-known effect see, e. g., reference \cite{UL}).
If one simplifies the form of states \ref{Fgstate1} and \ref{Fgstate2} 
by setting $|E_{0}\rangle = |J,m \rangle $ and
$|E_{0}\rangle = (|J,m \rangle +|J, m +1 \rangle)/ \sqrt 2$ in correspondence
to the appropriate values of $\Delta$,
the dimensionless mean AM per boson is found to be
$$
\langle l_z \rangle = \frac{1}{2}+\frac{m}{2J}, \quad
\langle l_z \rangle = \frac{1}{2}+\frac{m}{2J} \pm \frac{1}{4J},
$$
for $m -1/2 < \Delta/2|g| < m +1/2$ and $\Delta/2|g| = m \pm 1/2$,
respectively, corresponding to the two choices of the Fock ground state 
$|E_{0}\rangle$. This illustrates the AM step character (related to
the Hess-Fairbank effect) as well as its
"singular" behavior when $\Delta/2|g| = m \pm 1/2$. Notice that considering
the simplified form for $|E_{0}\rangle$ is equivalent to assume the net predominance of one or two components. The results just found are consistent
with the limit $\tau \to 0$, where $H_a = |g|J_3^2 -\Delta J_3$ 
can be diagonalized in a direct way.
\medskip

\noindent
\textbf{Josephson and Rabi regimes}.
In these cases $1/J^2 << \tau << 1$ and $1 << \tau$, respectively. 
Based on the above formulas, one finds
$\langle J_3 \rangle \simeq {{J} \tau \Delta  }/{ 2 V_0}$ (Josephson case)
and $\langle J_3 \rangle \simeq {J \Delta  }/{ 2 V_0}$ (Rabi case)
giving the mean dimensionless AM per boson 
$$
\langle l_z \rangle =
\frac{1}{2} \left [ 1 + \frac{\tau \Delta }{ 2V_0 } \right ] 
\, , \quad
\langle l_z \rangle = 
\frac{1}{2} \left [ 1+ \frac{ \Delta }{2 V_0 } \right ]\, ,
$$
respectively.
Owing to formulas \ref{J3} and \ref{J3vin}, in the Josephson case,
the range of parameter $\Delta$ is $[-2J|g|, \, \le 2J|g|]$.
For this regime, the further condition \ref{Herm} reduces to
$(2n \tau^{1/2}/J)^{1/2}  + (\Delta/ 2J|g|) < 1$.
In the Rabi case, condition \ref{J3vin}
on $\langle J_3 \rangle$ entails that $\Delta$ ranges in 
$[-2J \tau |g|, \, 2J \tau |g| ]$ ($2J \tau |g|= 2V_0$) 
which is, in principle, much larger than the range allowed in 
the Josephson case. Considering once more condition \ref{Herm}, 
this gives in the Rabi case $ (2n \tau^{1/2}/J)^{1/2}  +(\Delta/ 2V_0) < 1$.
On easily checks that weakly excited states $|E_n \rangle$ satisfy the
conditions on the restricted range of $\Delta$ provided
$n<<J$, and $|\Delta| << 2 J|g|$, $|\Delta| << 2V_0$ in the Josephson case
and in the Rabi case, respectively. In both cases the latter inequalities
represent condition \ref{J3vin} in its stronger version.

\section{The In\"{o}n\"{u} - Wigner contraction in the repulsive case}
\label{sez3}

%
The classical study of repulsive Hamiltonian
$H_r = -(|g| \, J_{3}^{2} +2V_0J_{1}+ \Delta J_{3})$,
discussed in \ref{appB},
shows that, with $\tau =V_0/J|g| >1$ (Rabi regime), 
the energy minimum is such that
$J_1 = J$, $J_2=J_3= 0$. As shown by equation \ref{min2},
a generic state near the minumun is such that 
$J_1 \simeq J$, $|J_2|$, $|J_3| \ll J$.
In the Fock/Josephson regimes, where $\tau =V_0/J|g| < 1$, 
Hamiltonian $H_r$ displays two minimum-energy states
(see equation \ref{min3}) entailing low-energy configurations
characterized by $J_3 \simeq \pm J$, $ |J_2|$, $|J_1| \ll J$.

\subsection{Repulsive regime with $\tau > 1$ }
 
In the Rabi regime ($\tau >1$), the CPA valid for the attractive model
can be implemented again. Then assuming $h_{1}$, $h_{2}$, $h_{3}$ 
as in formulas \ref{contr1}
the result of the contraction gives
$J_{1} \rightarrow J-n$, $J_{2}\rightarrow -\sqrt{J} p$,
and $J_{3}\rightarrow \sqrt{J} q$, which reduce $H_r$ to a quadratic form.
By defining $Q=q-c$, with $c= {\sqrt J} \Delta/(2V_0W^2)$, the final form of 
$H_a$ is found to be
\begin{equation}
H_a= V_0 \left [p^2 + W^2 Q^2- 2J -\frac{ {J} \Delta^2 }{4V^2_0 W^2} \right ].
\label{Hac2}
\end{equation}
Since the eigenvalues of $p^2+ W^2 q^{2}$ are $\Lambda_n = 2W(n+1/2)$, 
the spectrum of $H_r$ is
\begin{equation}
E_n = V_0 \left [2W(n+1/2) - 2J -\frac{ {J} \Delta^2 }{4V^2_0 W^2} \right ].
\label{Erep}
\end{equation}
As in the attractive case, the eigenfunctions $\Phi_n(Q)$ of Hamiltonian 
\ref{Hac2} allow one to determine components $X_m$ through the formula
$X_m (E_n) = \Phi_n(Q)$.
%
The energy eigenstates turn out to be
\begin{equation}
|E_{n} \rangle = \Sigma_m X_m (E_n ) |J; m \rangle \, , \quad  
X_m (E_n ) = N_{n}H_{n}({\sqrt W} Q) e^{-\frac{W Q^{2}}{2} },
\label{EERtsm}
\end{equation}
with $Q= {m}/{\sqrt J} -c$.
This description is valid if the conditions on 
the gaussian deviation and the Hermite-polynomyal zeros
${1}/{\sqrt J} < \sqrt{ {2}/{W} }$
and $|c| < {\sqrt J} -{\sqrt {{2n}/{W}} }$, respectively,
which can be rewritten as $1 < 2J \tau/(\tau -1)$ and 
$|\Delta| \tau /[2 V_0 (\tau-1)] 
< 1 -\{ 2n \tau^{1/2}/[J (\tau-1)^{1/2}] \}^{1/2}$,
are satisfied. For $\tau \gg 1$, the first condition is fulfilled, 
while the second one gives ${\Delta}/{2 V_0} < 1 -\sqrt {2n/J}$.
The latter is satisfied if ${\Delta}/{2 V_0} < 1$. Weakly excited
states $|E_n \rangle$ with $n>0$ can be also considered provided 
$J\gg 2n$. Under such conditions, the mean dimensionless AM per boson 
is a linear function of $\Delta$
$$
\left<l_{z}\right> =\frac{1}{2} (1 + \langle J_3 \rangle/J) =
\frac{1}{2} \left [ 1 + \frac{\Delta \tau}{2 V_0 (\tau-1)} \right ]
\, , \quad
\langle J_3 \rangle = c \sqrt J= {J \Delta \tau}/{2 V_0 (\tau-1)}\, ,
$$
giving $\left<l_{z}\right>
\simeq (1 + {\Delta}/{2 V_0 } )/2$ for $\tau >>1$. Notice that
$\langle J_3 \rangle$ coincides with formula \ref{minRrabi} for the
minimum of the classical repulsive model and that,
in the Rabi regime, $\left<l_{z}\right>$ has the same form 
both for attractive bosons ($g<0$) and for repulsive bosons ($g>0$).

\subsection{Repulsive case with $\tau < 1$ }

In this case, the classical ground-state configuration corresponds
to two minima. The contraction scheme can be implemented in two ways
by assuming $ h_{2} =xJ_2$, $h_{1} =x J_1$, and $h_{3} =J_3 \mp {I}/{x^{2}}$  
which entails
$[h_1, h_2]= \pm ix^2h_3$,
$[h_2, h_3]= ih_1$, and
$[h_3, h_1]= ih_1$.
Notice that $h_{3} =J_3 \mp {I}/{x^{2}}$ allows to describe the
two classical minima by further selecting a suitable definition
for $h_3$.  The result of the contractions demonstrates the two
possible choices
\begin{equation}
h_{3} = -n, \, h_{2} = J_2/\sqrt{J} \rightarrow  p, \,
h_{1} = J_1/\sqrt{J} \rightarrow q,
\label{cont5}
\end{equation}
and
\begin{equation}
h_{3} = +n, \, h_{2} = J_2/\sqrt{J} \rightarrow  -p, \,
h_{1} = J_1/\sqrt{J} \rightarrow q,
\label{cont6}
\end{equation}
that are naturally associated to the $J_3$-positive 
and $J_3$-negative minimum, respectively.
The repulsive Hamiltonian
$H_r = - ( |g|\, J_{3}^{2} + 2V_0J_{1}+ \Delta J_{3} )$
thus can be cast in the two (local) forms
\begin{equation}
H_r= -|g| \Bigl [  J^2-2J\, n + 2\tau J^{3/2} q 
+\frac{s \Delta}{|g|}(J- n) \Bigr ],
\label{RattH}
\end{equation}
where $s= \pm$ recalls the presence of two minima.
Notice that $H_r$ could be diagonalized by means of the procedure
used in the attractive case, provided one adopts the rotated basis
$\{ |m \rangle_1 = \exp(-i\pi J_2 /2) |m \rangle \}$ of $J_1$
and regards the $q$ eigenvalues
$m/\sqrt J$ as a continuous index. Unfortunately, while the evaluation
of the energy eigenvalues is very easy in the "rotated" $J_1$ basis 
$\{ |m \rangle_1 = \exp (-i\pi J_2/2) |m \rangle, |m| \le J \}$, 
the eigenstates must be counter-rotated to recover the 
$J_3$-basis representation that we have adopted in the
other cases/regimes. This is a difficult problem in that recovering
the eigenstates description in the $J_3$ basis requires that the
transformation matrix element
$\langle m'| \exp (-i\pi J_2/2)|m \rangle$ is calculated explicitly
and is formulated in the limit where $m/\sqrt J$ is a continuous index.

To skip this problem, we observe that, owing to formulas \ref{cont5} and 
\ref{cont6} derived by the contraction procedure, 
$J^2_3 +J^2_2 +J^2_1 = J(J+1) \simeq J^2$ can be rewritten as 
$J^2_3 \simeq 2J n -J^2$ while $J_3 = \pm (J-n)$.
We thus obtain the linearized expression $J^2_3 \simeq -J^2 \pm 2J J_3$.
Hamiltonian \ref{RattH} reduces to
$H_r= -|g| \bigl [ -J^2 \pm 2J J_3 + 2\tau J J_1 
+ ({\Delta}/{|g|}) J_3 \bigr ]$,
whose digonalization is rather simple owing to the linear dependence 
on su(2) generators. Rewriting the latter as
$H^{\pm}_r= 
|g| \Bigl [ J^2 \mp (2J \pm \delta) J_3 - 2\tau J J_1  \Bigr ]$
where $\delta =\Delta/|g|$,
the unitary transformations $U_{\pm}= \exp (\mp i J_2 \phi_{\pm} )$
entail
\begin{equation}
H^{\pm}_r = 
|g| \Bigl [ J^2 \mp R_{\pm } \, U_{\pm } J_3 U^{+}_{\pm }
 \Bigr ]\, ,
\label{RaH3}
\end{equation}
with $R_{\pm } = \sqrt{(2J \pm \delta)^2+4 \tau^2 J^2}$.
The action of $U_{\pm}$ is given by
\begin{equation}
U_{-} J_3 U^{\dagger}_{- }
= J_3  \cos \phi_{-} - J_1 \sin \phi_{-} \, , 
\quad
U_{+ } J_3 U^{\dagger}_{+ }
= J_3 \cos \phi_{+} + J_1 \sin \phi_{+} \, ,
\quad
\label{rot1}
\end{equation}
where
angles $\phi_{\pm}$ are definded by 
${\rm tg} \phi_{-} = {2\tau J}/{(2J - \delta)}$,
${\rm tg} \phi_{+} = {2\tau J}/{(2J +\delta)}$.
The energy spectrum is thus represented by
the eigenstates and the eigenvalues
\begin{equation}
|E^{\pm}_m \rangle = U_{\pm } \, |m \rangle ,
\quad
E^{\pm}_m = 
|g| \Bigl [ J^2 \mp m \sqrt{(2J \pm \delta)^2+4 \tau^2 J^2 }  \Bigr ],
\label{VSp3}
\end{equation}
respectively.
One should recall that, within the present approximation scheme, these
eigenvalues are significant for $|m| \approx J$. Moreover, we notice that
$U_{\pm } \to \bf 1$ for $\tau\to 0$ thus reproducing the correct spectrum
of the uncoupled model.   
The eigenvalues corresponding to the
energy minima are obtained by setting $m =-J$ and $m =+J$ 
for $H^{-}_r$ and $H^{+}_r$, respectively, and read 
\begin{equation}
E^{\pm}_{M} (\delta) := E^{\pm}_{\pm J} = 
|g| \Bigl [ J^2 -J \sqrt{(2J \pm \delta)^2+4 \tau^2 } \,   \Bigr ].
\label{Egs}
\end{equation}
The choice of the signs $\pm$, and thus the recognition of
the lowest-energy states, is related to the sign of $\delta$.
This is discussed below.
%
%
\begin{figure}
 \begin{center} 
 \begin{tabular}{cc}
 \includegraphics[width=6.5cm]{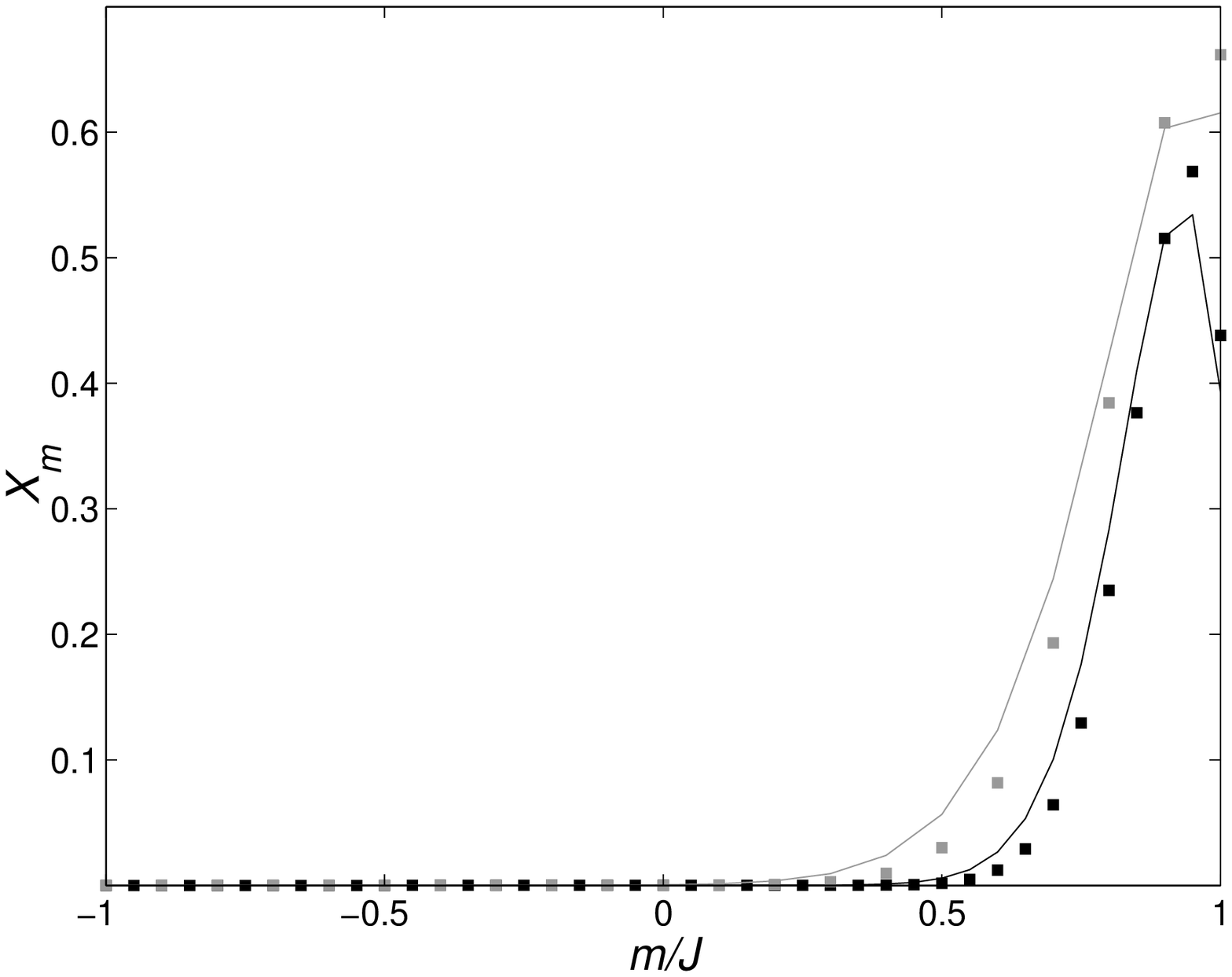}&
 \includegraphics[width=6.5cm]{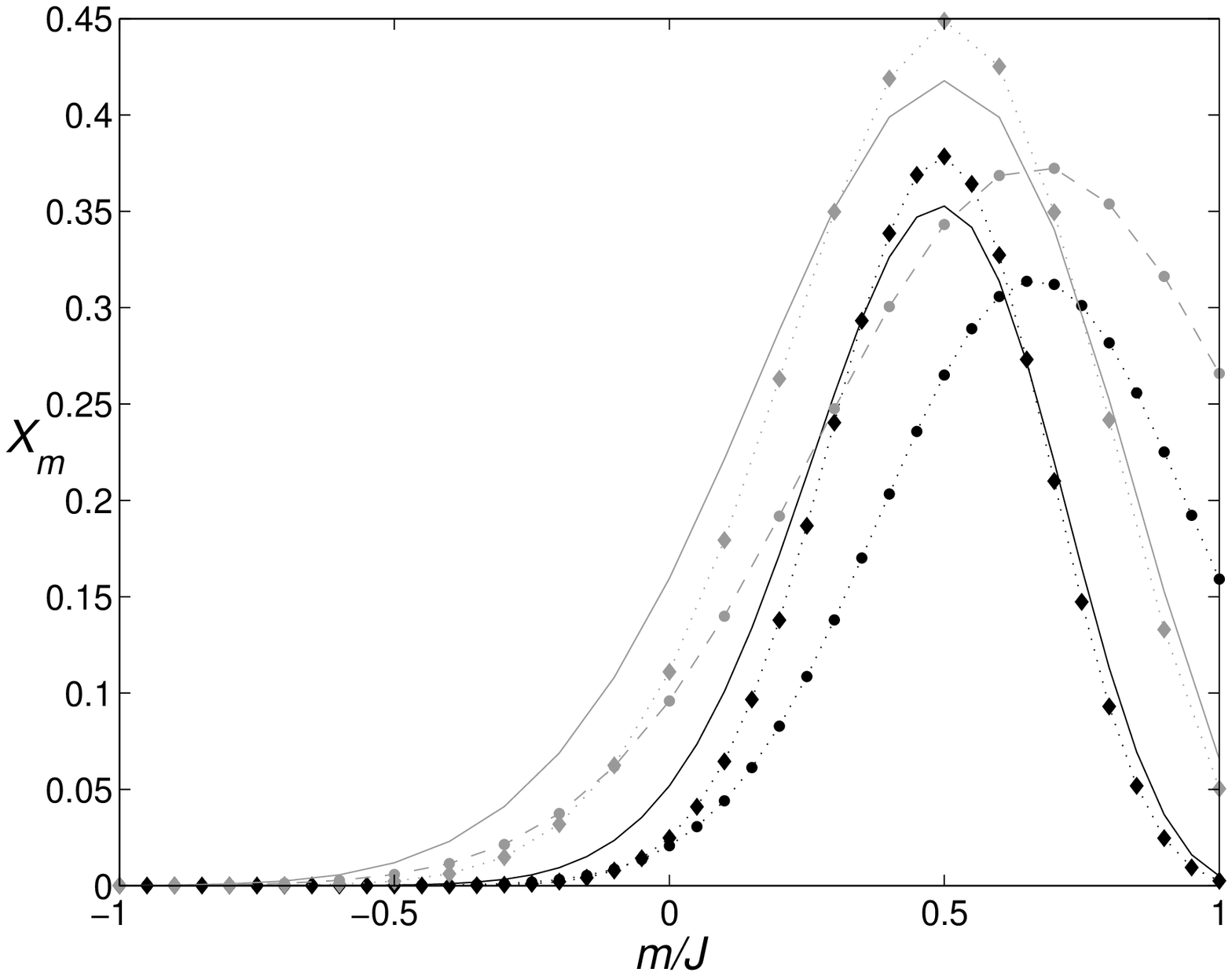}
 \end{tabular}
 \caption{\label{figura45} 
Both panels concern the repulsive case.
Grey (dark) squares, diamonds, points and (piecewise-linear,
continuous, dotted or dashed) curves are relevant to $N=20$ ($N=40$).
{\bf Left panel}: $\tau=0.6$ (Josephson regime), $\nu =0.8$.
Within the CMA, the ground-state components $X_m$'s, given by
formula \ref{GSt1} and described by squares, well approximate the $X_m$'s
(edges of the grey/dark piecewise-linear curves) calculated numerically.
{\bf Right panel}: $\tau=1.6$ (Rabi regime), $\nu =0.8$.
Points (diamonds) --joined by dashed/dotted lines to better distiguish different
cases-- describe ground-state $X_m$'s within the CMA (CSSA) referred to formula
\ref{EERtsm} (formula \ref{CSSA}).
Continuous piecewise-linear curves represent $X_m$'s obtained numerically. 
The CSSA is qualitatively better than the CMA approximation where curves are shifted on the right. Further comments are given in section \ref{conc}.
}
\end{center}
\end{figure}
%
The states associated with eigenvalues \ref{Egs}
take the form of su(2) coherent states \cite{ZFG}.
The standard su(2) picture of such states, also known as Bloch states,
is given by
\begin{equation}
|{- \, J}, \xi \rangle = e^{\xi J_+ - \xi^* J_-} |-\! J \rangle =
\sum^{2J}_{s=0} \frac{C_{Js} z^s |s-J \rangle }{(1+|z|^2 )^J}  
\label{CS1}
\end{equation}
with $C_{Js} = \sqrt {(2J)! /s! (2J-s)!}$, while the coherent-state labels
$z = |z| e^{i \theta}$ and $\xi = |\xi| e^{i \theta}$ are such that
$|z| = {\rm tg} |\xi|$, $z \in \bf C$. 
Since the minimum-energy states have the form
\begin{equation}
|E^{\pm}_{M} \rangle =
e^{\mp i J_2 \phi_{\pm} } \, |\pm J \rangle ,
\label{VSp4}
\end{equation}
where $\mp i J_2 \phi_{\pm} =\mp (\phi_{\pm}/2)(J_+ -J_-) $, the
link with the coherent-state picture is almost immediate. 
Upon setting $\xi = \mp \phi_{\pm}/2$, the corresponding $z$ reads 
$z= \mp {\rm tg}(\phi_{\pm}/2) = \mp {2\tau J}/({2J \pm \delta})$.
In view of this, eigenstate $|E^{-}_{M} \rangle$
takes the new form
\begin{equation}
|E^{-}_{M} \rangle = 
\cos^{2J} (\phi_-/2)
\sum^{2J}_{s=0} C_{Js} {\rm tg}^s(\phi_-/2) |s-J \rangle .
\label{GSt1}
\end{equation}
If $\delta <0$, state $|E^{-}_{M} (\delta) \rangle$ (we make
explicit the dependence from $\delta$ to illustrate clearly the
difference between the absolute minumum and the local minimum)
corresponds to the lowest-energy
state with eigenvalue
$E^{-}_{M} (\delta) = 
|g| [\, J^2 -J \sqrt{(2J +|\delta| )^2+4 \tau^2 } \, ]$,
since $E^{-}_{M} (\delta)< E^{+}_{M} (\delta)$ (see equation \ref{Egs}). 
The remaining state 
$|E^{+}_{M} (\delta) \rangle$ represents the local minimum
found in the classical dynamics.
In the opposite case $\delta > 0$, the lowest energy state 
identifies with $|E^{+}_{M} \rangle $. This in fact
corresponds to (see equation \ref{Egs})
$E^{+}_{M} (\delta) = 
|g| [\, J^2 -J \sqrt{(2J + \delta )^2+4 \tau^2 } \, ]$,
which satisfies $ E^{+}_{M} (\delta) < E^{-}_{M} (\delta)$
for $\delta>0$. Notice that
$E^{-}_{M} (-|\delta|) \equiv E^{+}_{M} (\delta)$. This feature
is important because it confirms the symmetry property 
$e^{+i \pi J_1} H_r (\delta) \, e^{-i \pi J_1}$ $=H_r (-\delta)$
of repulsive Hamiltonian
$H_r (\delta) = -|g| (J_3^2+ 2J\tau J_1 +\delta J_3)$
stating that the spectra of the cases $\delta>0$ and
$\delta<0$ must coincide, the relevant Hamiltonians being
related by a unitary transformation.
Based on this fact, we find as well
$|E^{+}_{M} (\delta) \rangle =
e^{+i \pi J_1} |E^{-}_{M} (-|\delta|) \rangle$.
By acting with $e^{i \pi J_1}$ on $|E^{-}_{G} (-|\delta|) \rangle$
we get the expression
\begin{equation}
|E^{+}_{M} (\delta) \rangle =
e^{-i J_2 \phi_{-} } e^{+i \pi J_1} \, |- \! J \rangle
= e^{i J \pi } e^{-i J_2 \phi_{-} } |+ \! J \rangle
\label{GSt2}
\end{equation}
[notice that $\phi_{-} = \phi_{-}(-|\delta|)$],
where we have used the property
of the $J_3$-basis states
$e^{i J_1 \pi } |m \rangle = e^{i J \pi } |-m \rangle$.
Upon observing that $\phi_{-}(-|\delta|)= \phi_{+}(+|\delta|)$
we conclude that the unitary transformation
reproduces, up to a phase factor, 
the diagonalization-process formula
$|E^{+}_{M} \rangle = e^{- i J_2 \phi_{+} } \, |+ \! J \rangle$
in a consistent way.
%
%
\begin{figure}[h]
\centering
\includegraphics[width=8.5cm]{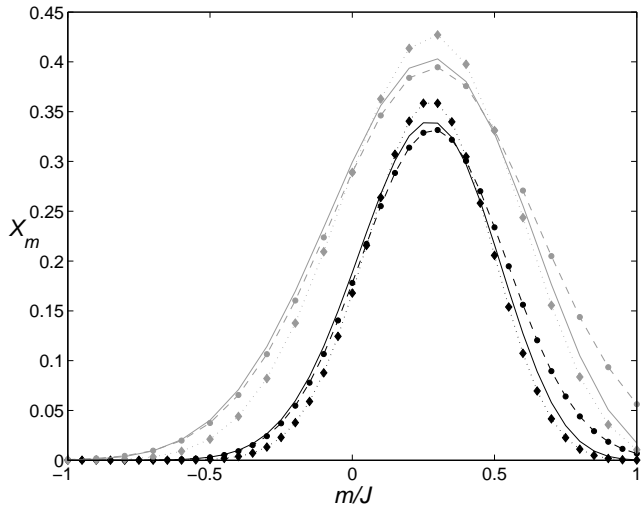}
\caption{Repulsive case, $\tau=2.4$ (Rabi regime), $\nu =0.8$.
Grey (dark) diamonds, points and piecewise-linear curves are relevant
to $N=20$ ($N=40$).
Diamonds (points) --joined by dashed/dotted lines 
to better distiguish different
cases-- describe ground-state $X_m$'s given by formula \ref{EERtsm}
(formula \ref{CSSA}) within the CMA (CSSA). Piecewise-linear curves have the usual meaning. Both CSSA and CMA are satisfactory. Further comments are given 
in section \ref{conc}.}
\label{figura6}
\end{figure}
%
%
%
Therefore, the ground state 
of the case $\delta >0$ is obtained
by calculating formula \ref{GSt2} explicitly, which
gives 
\begin{equation}
|E^{+}_{M} \rangle =
\cos^{2J} \left ({\phi_{+}}/{2} \right ) \,
\sum^{2J}_{s=0} 
{C_{Js} {\rm tg}^s \left({\phi_{+}}/{2} \right ) \, |J-s \rangle }  ,
\label{GSt3}
\end{equation}
where $\phi_{+}(+|\delta|)= \phi_{-}(-|\delta|)$. We notice that
$|E^{+}_{M} \rangle$ corresponds to a coherent state 
$|+ \! J, \xi \rangle = e^{\xi J_+ - \xi^* J_-} |+ \! J \rangle$
whose extremal state is $|J \rangle$ (instead of $|-J \rangle$)
where $|v|=tg|\xi|$ with $v = -tg(\phi_{-}/2)$ reproduces \ref{GSt3}.
As in the case $\delta <0$, the remaining state
$|E^{-}_{M} (\delta) \rangle$ describes
the quantum counterpart of the local minimum.
The expectation value of $J_3$ is easily carried out. 
By using equations \ref{rot1}, one finds
($\langle J_k \rangle_{\pm} =
\langle E^{\pm}_{M} | J_k |E^{\pm}_{M} \rangle$, $k=1,2,3$)
$\langle J_3 \rangle_{\pm } =
\langle \pm J | ( J_3 \cos \phi_{\pm}  
\mp J_1 \sin \phi_{\pm} ) \, |\pm J \rangle$,
$\langle J_1 \rangle_{\pm } = J \sin \phi_{\pm}$,
$\langle J_2 \rangle_{\pm} =0$,
namely 
\begin{equation}
\langle J_3 \rangle_{\pm}= 
\frac{\pm J}{\sqrt{1+ \mu_{\pm}^2 }} , \quad
\langle J_1 \rangle_{\pm}= 
\frac{ J \mu_{\pm} }{\sqrt{1+\mu_{\pm}^2}} ,
\label{J3MV}
\end{equation}
where $\mu_{\pm} = 2\tau J/(2J \pm \delta)$,
which, expanded up to second order in $\tau$, appear to be consistent
with the classical values \ref{dmin} of the minimum-energy 
configurations. The choice $+$ ($-$) for
the lowest-energy state, corresponding to $\delta >0$ ($\delta <0$),
entails $2J \pm \delta = 2J +|\delta|$ in $\mu_{\pm}$. 
Thus $\langle J_3 \rangle_{+}$ and $\langle J_3 \rangle_{-}$ simply differ
of a factor $-1$. 
In passing we notice that states $|E^{\pm}_{M} (\delta) \rangle$ with the same
$\delta$, should satisfy the condition $\langle E^{+}_{M} |E^{-}_{M} \rangle =0$
they corresponding to different eigenvalues.
$|E^{\pm}_{M} (\delta) \rangle$ obtained within the CPA can be shown to be
almost orthogonal \cite{ort}.
Excited states labeled by $m=\pm (n-J)$ with $n<<J$ can be derived
explicitly from formula \ref{VSp3}. By expressing them as
$|E^{\pm}_m \rangle = U_{\pm } J^n_{\mp}|\pm J \rangle / ({n! C_{Jn}})$,
one obtains
$
|E^{\pm}_m \rangle = 
( U_{\pm } J_1 U^{\dagger}_{\pm }  \mp i J_2)^n 
|E^{\pm}_{\pm J} \rangle/ ({n! C_{Jn}}) \, ,
$
with $U_{\pm } J_1 U^{\dagger}_{\pm }
= \cos (\phi_{\pm}) J_1 \mp \sin (\phi_{\pm}) J_3 $,
that can be used to calculate the expectation values of
operators $J_k$, $k= 1,2,3$ .
The condition under which
the eigenvalue that corresponds to the local minimum represents the first
excited state can be determined quite easily (e. g., for $\delta <0$) 
from $ E^{+}_{M} (\delta) \le E^{-}_{m} (\delta)$ with $m= -J+1$.

Within Fock and Josephson regimes ($\tau < 1$),
the AM per boson is readily evaluated from formula \ref{J3MV} giving
$
\langle \ell_z \rangle = [ 1 \pm 
(2J \pm \delta )/{\sqrt{4\tau^2 J^2+ (2J \pm \delta)^2}} ] /2
$.
If $\tau << 1$, due to
$\phi_{\pm} \simeq 2\tau J/(2J \pm \delta)$ 
and in view of equations \ref{GSt1} and \ref{GSt2}, 
the ground state reduces to
$
%
|E^{\mp}_{G} \rangle \simeq
[1 -2J (\phi_{\mp}/2)^2] \bigl [ |\mp J \rangle + 
{\sqrt {{J}/{2} }} \phi_{\mp} |\mp J \pm 1 \rangle \bigr ]
$
where $-$ and $+$ are related to the cases $\delta <0$ and 
$\delta >0$, respectively. 
Thus in the Fock regime ($\tau << 1/J^2$) it is natural to set
$\phi_{\pm} \simeq 0$. By neglecting also the first order
corrections, the ground state is approximated by
$|E_{G} (\delta) \rangle =  \theta (\delta) |J \rangle + 
\theta (-\delta) |- J \rangle$
which, inserted in formula \ref{J3MV}, gives 
$\langle \ell_z \rangle =  \theta (\delta) = (1\mp 1)/2$.
This well matches the case $\tau =0$ where 
$E_{G} (\pm |\delta|) = -(|g| J^2 \pm J \Delta)$ with $\delta=\Delta/|g|$.
%

\section{The coherent-state semiclassical approximation.}
\label{sez4}
An alternative way to approximate both the ground state and the corresponding
energy is to find the quantum counterpart of a classical configuration in terms of coherent states. If the hamiltonian algebra of a given model is 
known together with the coherent state relevant to such an algebra, classical
variables can be put in a one-to-one correspondence with the complex labels
parametrizing a coherent state \cite{ZFG}. 
This is the case for Hamiltonian \ref{Hatt} 
and \ref{Hrep} that are written in terms of su(2) generators $J_3$, $J_{\pm}$.
Coherent states $|{- \, J}, \xi \rangle$ of algebra su(2) are defined by
equation \ref{CS1}. The latter allows one to parametrize a coherent state
by $z$ since $\xi = |\xi| e^{i \theta}$ is related to 
$z = |z| e^{i \theta}$ by $|z| = {\rm tg} |\xi|$. For
a generic $| z \rangle$ the expectation values 
$\langle J_k \rangle = \langle z | J_k | z\rangle$, $k= \pm ,3$, given by
\begin{equation}
\langle J_3 \rangle = 
J (|z|^2-1)/(|z|^2+1 ) , \quad
\langle J_+ \rangle = 2J z^*/(|z|^2+1 ) , 
\label{CSSA}
\end{equation}
with $\langle J_- \rangle =\langle J_+ \rangle^*$,
allow one to determine $z$ when $\langle J_k \rangle$ are known. Notice that
$\langle J_1 \rangle= (\langle J_+ \rangle +\langle J_- \rangle)/2$ and
$\langle J_2 \rangle= (\langle J_+ \rangle -\langle J_- \rangle)/2i$. 
Therefore classical configurations characterized by known values of 
$J_1$, $J_2$ and $J_3$ can be associated with a specific $z$ by
identifying each classical $J_k$ with $\langle J_k \rangle$ and
observing that, owing to equations \ref{CS1}, the phase $\theta$ of $z$ 
coincides with the phase of $J_+ = J_1+iJ_2$ while $|z|^2 = (J+J_3)/(J-J_3)$.
Recalling that
this assumption becomes exact in the semiclassical limit $J \to \infty$,
we name the map $J_1, J_2, J_3 \to z$
coherent-state semiclassical approximation (CSSA). 
Determining $J_k$'s that characterize the classical energy minimum thus
provide the ground-state approximation $| E_M \rangle \simeq | z \rangle$
where $| z \rangle$ is determined by the previous semiclassical map.
The correponding energy is obtained by $E^{sc}_M =\langle z | H | z \rangle$.

\section{Conclusions}
\label{conc}

We have discussed the effectiveness of the CPA
based on the In\"on\"u-Wigner transformation by comparing the ground state
(GS) obtained in the various regimes of both the repulsive and the attractive models with the exact lowest-energy eigenstate determined numerically. In the
attractive case ($g<0$), both for $\tau <1$ and for $\tau >1$, and
in the repulsive case ($g>0$) for $\tau > 1$ the CPA leads to approximate
$X_m$'s of weakly excited states through the eigenfunctions of
equivalent harmonic-oscillator problems represented by formulas
\ref{EEAtlarge} and \ref{EERtsm}, respectively. Due to the presence of two
classical minima in model \ref{Hrep},
the repulsive case with $\tau < 1$ requires that a different diagonalization scheme is developed after implementing the CPA on Hamiltonian \ref{Hrep}. 
This involves weakly excited states represented in terms su(2) coherent
states \ref{GSt1} and \ref{GSt3}.
%
%
%
\begin{figure}[h]
\centering
\includegraphics[width=7.3cm]{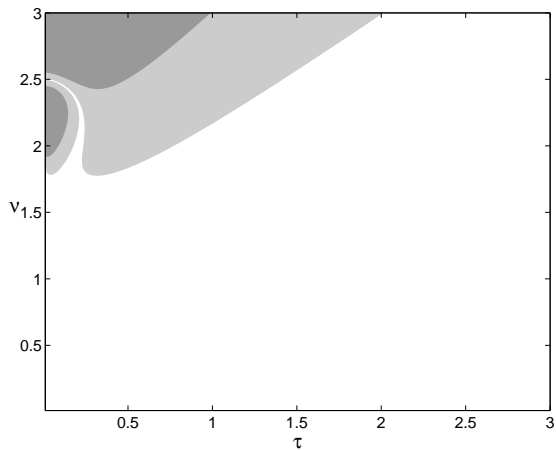}
\caption{Attractive case. Comparison of the exact 
ground-state (GS) energy with the ground-state energy \ref{EAval} 
within CPA. Different
colors in the $\tau \nu$ plane are related to different value of indicator
$\sigma$. White regions are characterized by an excellent agreement of the exact and the approximated GS energies. See
section \ref{conc} for details.}
\label{figura7}
\end{figure}
%
%
In the attractive case, figures \ref{figura12} show
that the exact components (calculated numerically) are almost
indistinguishable from components $X_m$'s obtained within the CPA and 
described by formula \ref{EEAtlarge}. The cases $N=20$ and $N=40$ that correspond to $\tau = 0.02 >1/J^2 = 0.01$ and $\tau = 0.02 >1/J^2 = 0.0025$, respectively, describe the approach from above to the lower bound of Josephson regime.
In the repulsive case, figures \ref{figura45} allow one to compare 
the exact components (calculated numerically) with components $X_m$'s
obtained within the CPA and described by formula \ref{GSt1}
for $\tau = 0.6$ (Josephson regime), $\nu =0.8$ and $N= 20, \, 40$,
and by formula \ref{EERtsm}
for $\tau = 1.6$ (Rabi regime), $\nu =0.8$ and $N= 20, \, 40$.
While in the first case formula \ref{GSt1}, representing a 
su(2) coherent state, provides a satisfactory approximation, 
in the second case formula \ref{EERtsm}
exhibits a shift on the right of highest weight components 
$X_m$'s that, in addition, are smaller than the exact ones. 
In figure \ref{figura45}
(right panel) $X_m$'s evaluated within the CSSA better match the exact ones 
both qualitatively and quantitatively.
When $\tau$ is increased (see figure \ref{figura6}),
the CPA approximation (CSSA) is satisfactory even if it tends to underestimate (overestimate) exact $X_m$'s.
%
\begin{figure}
 \begin{center}
 \begin{tabular}{cc}
 \includegraphics[width=6.5cm]{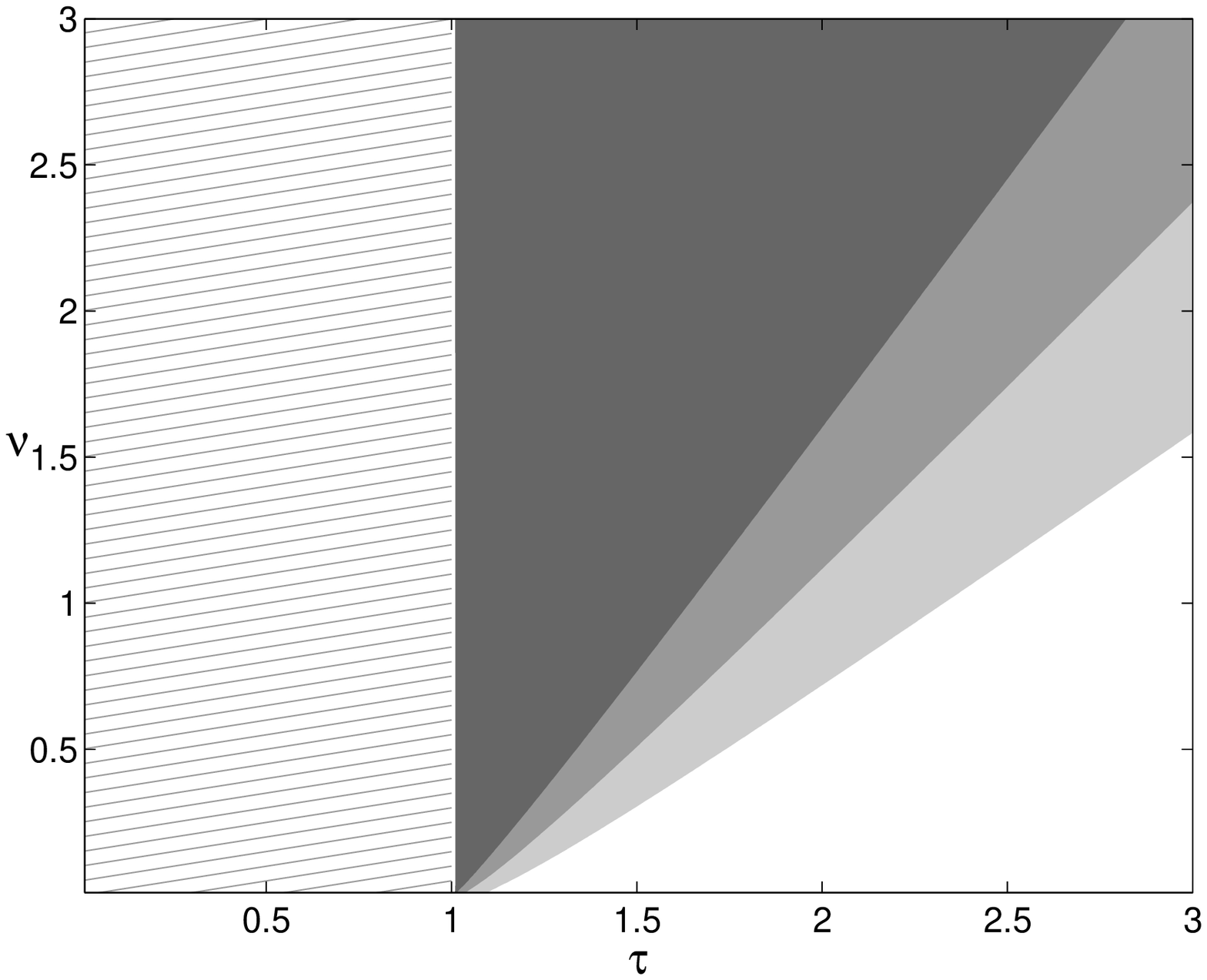}&
 \includegraphics[width=6.5cm]{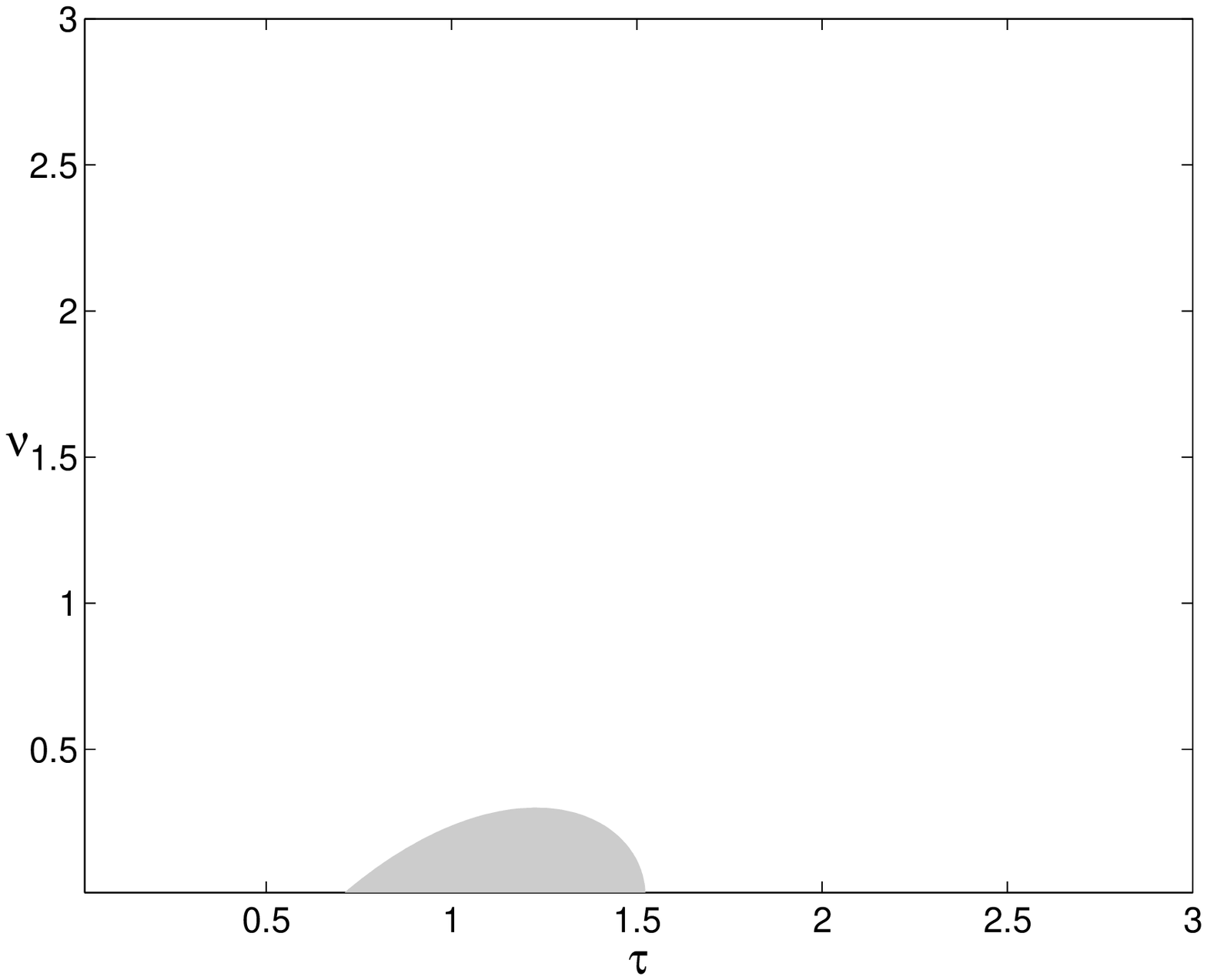}
 \end{tabular}
 \caption{\label{figura89} 
Repulsive case. Comparison of the exact ground-state (GS) energy
with approximate GS energies. Different
colors in the $\tau \nu$ plane are related to different value of indicator
$\sigma$. White regions are characterized by an excellent agreement of the exact and the approximated GS energies. Details are discussed in section \ref{conc}.
{\bf Left panel}: $\sigma$ for the GS energy \ref{Erep} ($\tau> 1$).
{\bf Right panel}: $\sigma$ for the GS energy within the CSSA.
(expectation value of $H_r$ for the GS relevant to formula \ref{CSSA}).}
\end{center}
\end{figure}
%
Figures \ref{figura7} and \ref{figura89} illustrate, through the
parameter $\sigma = (E^e_{M} - E^{ap}_{M})/\Delta E $, the deviation
of the GS energies obtained within the CPA or the CSSA from 
the GS energy calculated numerically. Energies $E^e_{M}$, $E^{ap}_{M}$, and 
$\Delta E$ are the exact GS energy, the approximated GS energy and the energy range defined as $\Delta E =E^e_{max} -E^{e}_{M}$, respectively. 
$E^e_{max}$ is the exact maximum energy. White, light grey,
and dark grey colors identify the regions in the 
$\tau \, \nu$ plane where $\sigma < 0.001 $, $0.001 < \sigma < 0.01 $
and $0.01 < \sigma < 0.1 $, respectively. In figure \ref{figura7},
describing the attractive case, $E^{ap}_{M}$ is given by formula \ref{EAval}.
$E^{ap}_{M}$ well approximates the exact GS energy in the large
(white) region in the $\tau \, \nu$ plane. The repulsive case is considered
in figure \ref{figura89}. In the left panel, $E^{ap}_{M}$ given by formula 
\ref{Erep} is shown to well approximate the exact GS energy in a
rather restricted region in the $\tau \, \nu$ plane. On the contrary, right
panel shows that evaluating $E^{ap}_{M}$ based on ground-state \ref{CSSA}
within the CSSA provides the best approximation ($\sigma < 0.001 $) almost everywhere.
Concluding, except for the repulsive Josephson regime, where the CPA is not
satisfactory, both the CPA and the CSSA provide a satisfactory approximation.
The CPA is particularly good in the attractive-boson case. Among the many
applications to bosonic-well systems currently studied, such approaches
seem quite appropriate for studying the low-energy spectrum of the three-well boson systems where the complexity of the energy-level structure mirrors the
dynamical instabilities of the chaotic three-well classical dynamics
\cite{prl90}. The study of similar aspects in the three-AM mode rotational
fluid outlined in \cite{UL} is currently in progress.
%

\begin{appendix}

\section{Classical energy minima}
\label{appB}

The classical version of the attractive model \ref{Hatt} 
displays a dynamics characterized by four (two) fixed points if $1 \gg \tau$ 
($\tau \gg 1$). This can be seen by considering the relevant motion 
equations
\begin{equation}
{\dot J}_1 =  (\Delta -2|g| J_3) J_2 , \quad
{\dot J}_3 =  -2 V_0 J_2 , \quad
{\dot J}_2 =  2(|g| J_1 + V_0) J_3 - \Delta J_1 ,
\quad
\label{HEatt}
\end{equation}
equipped with the motion constant
$J^2 = J_3^2+ J_2^2 +J_1^2$, that
entail the fixed-point equations
$J_2=0$, $2|g| J_3J_1 + 2V_0 J_3 -\Delta J_1 =0$,
with the constraint $J^2 = J_3^2 +J_1^2$.
Their exact solution involves a fourth-order equation
in $J_3$, except for $\Delta =0$ 
when the possible solutions are either 
$J_1 = -V_0/|g|= -J \tau$ or $J_3 =0$. 
In the general case $\Delta \ne 0$, if $1 \gg \tau $ and $J|g| > \Delta > 0$
(namely, for $\Delta$ sufficiently small), 
the searched solutions are such that either $J_3 \simeq \pm J$,
$J \gg |J_1|$, or 
\begin{equation}
J_1 \simeq \pm J, \quad J \gg J_3>0 \, .
\label{min1}
\end{equation}
This feature can be proved explicitly. Particularly, the 
second pair of solution is obtained
by implementing the approximation 
$J_1 = s {\sqrt {J^2 - J_3^2}}$ $\simeq s J(1- J_3^2/2J^2)$, $s=\pm 1$.
Neglecting the third order terms in $J_3/J$,
the second fixed-point equation becomes
$({\Delta}/{2J}) J_3^2 + 2J|g| (1+s\tau) J_3 -J \Delta = 0$,
whose roots are found to be
$J_3 = 2 J \sigma^{-1}_s [-1 \pm {\sqrt { 1 + \sigma_s^2/2  } } ]$ with
$\sigma_s = \Delta/[J|g|(1+s\tau)]$.
While the negative root must be discarded because it
entails $|J_3|> J$, the positive root --this can be shown to 
describe both a minimum ($s=+1$) and 
a saddle point ($s=-1$)-- can be
approximated as
\begin{equation}
J_3 \simeq \frac{J \tau \Delta}{2V_0 (1 +s \tau)},
\label{minR}
\end{equation}
if $\delta=\Delta/|g| < J $.
When $\tau > 1$ (and thus 
for $\tau \gg 1$) the choices $s=-1,\, +1$ are related to a 
maximum and a minimum, respectively.
Notice that the previous formula giving the $J_3$ coordinate
is well defined for the minimum ($s=+1$) also when $\tau \gg 1$.

Let us consider now the (classical) repulsive model \ref{Hrep}.
The corresponding Hamiltonian equations read
\begin{equation}
{\dot J}_1 =  (\Delta +2|g| J_3) J_2 ,
\quad
{\dot J}_3 =  -2 V_0 J_2 ,
\quad
{\dot J}_2 =  2(V_0 -|g| J_1 ) J_3 - \Delta J_1 ,
\quad
\label{HEatt}
\end{equation}
and exhibit once more the motion constant $J^2 = J_1^2 + +J_2^2 + J_3^2$.
For $\Delta =0$ and $\tau >1$, 
the energy minimum is easily shown to correspond to
$J_1 = J$, $J_2 = J_3= 0$. 
Thus a generic state near the minumun is such that  
\begin{equation}
J_1 \simeq J, \quad |J_2|, \, |J_3| \ll J \, .
\label{min2}
\end{equation}
If $\Delta \ne 0$, provided
$\Delta/J|g|$ is sufficiently small, this statement is certainly valid 
for $1 \ll \tau =V_0/J|g|$ (Rabi regime). 
In fact, by setting 
$J_1 = {\sqrt {J^2 - J_3^2}}$ $\simeq  J(1- J_3^2/2J^2)$ and
neglecting the third order terms in $J_3/J$
in the fixed-point equation
$0= 2(V_0 -|g| J_1 ) J_3 - \Delta J_1$, one finds
$({\delta}/{2J}) J_3^2 + 2J (\tau-1) J_3 -\delta J = 0$,
whose roots are found to be
$J_3 = 2J\alpha^{-1} [-1 \pm {\sqrt {1 + \alpha^2/2}} ]$, with
$\alpha= \Delta/[J |g| (\tau-1)]$. Discarding the negative root which
entails $|J_3|> J$, the positive root 
can be approximated as
\begin{equation}
J_3 \simeq \frac{\Delta}{2|g| (\tau -1)}
=\frac{J \tau \Delta}{2V_0 (\tau -1)},
\label{minRrabi}
\end{equation}
if $\Delta/ J |g| \ll \tau -1$. 
In the Rabi regime where $1 \ll \tau \simeq \tau -1$ such
condition reduces to $\Delta \ll V_0 $. 
In the Fock/Josephson regimes, where $\tau < 1$, 
the two configurations 
$J_1 = \tau J$, $J_3= \pm J{\sqrt {1-\tau^2}}$
are found to minimize the energy if $\Delta =0$.
This suggests that, even with $\Delta \ne 0$, 
low-energy states are such that
\begin{equation}
J_3 \simeq \pm J, \quad |J_2|, \quad |J_1| \ll J \, .
\label{min3}
\end{equation} 
To obtain the energy-minimum configurations,
in addition to $J_2 =0$, we consider
the second fixed-point equation under the approximation
$J_3= s {\sqrt { J^2-J_1^2}} \simeq s J(1 -J_1^2/2J^2)$
with $s =\pm 1$.
Neglecting the third order terms in $J_1/J$,
the latter entails
$0= ({V_0}/{J}) J_1^2 + (2|g| J - s \Delta ) J_1 -2V_0 J$,
which supply, with $s=+1$, two minimum-energy configurations
($\delta =\Delta/|g|$)
\begin{equation}
J_1 \simeq \frac{\tau J}{1+ s \delta/ 2J}, \quad
J_3 = s J{\sqrt { 1-(J_1/J)^2}}
\simeq s J \Bigl [ 1 - \frac{2J^2 \tau^2}{(2J+s \delta)^2} \Bigr ]\, .
\label{dmin}
\end{equation}
These reproduce correctly the formula of the case $\Delta =0$.

\end{appendix}



\begin{thebibliography}{99}
\bibitem{ort}
An explicit calculation gives
$\langle E^{+}_{M} |E^{-}_{M} \rangle= [{\nu^2 }/{(1 +\nu^2)} ]^{J}$, 
with $\nu={\rm tg} [(\phi_+ +\phi_-)/2]$.
This term certainly vanishes being ${\nu^2 }/{(1 +\nu^2)} <1$ and $2J=N >>1$.
Since ${\nu^2 }/{(1 +|\nu|^2)} \simeq \tau^2 [{2J^2}/{(4J^2-\delta^2)}] $
and $\tau < 1$ then $\langle E^{+}_{M} |E^{-}_{M} \rangle \to 0$ very rapidly due to the factor $\tau^{2J}$.
\end{thebibliography}

\end{document}